\documentclass[a4paper,twocolumn,numberedappendix,twocolappendix,iop]{openjournal}
\usepackage{graphicx,amsmath,amssymb,amstext}
\usepackage[usenames,dvipsnames]{xcolor}
\usepackage{amsbsy,amsfonts,amsthm,bm}
\usepackage{ulem}
\usepackage[colorlinks,linkcolor=blue,citecolor=blue,urlcolor=blue ]{hyperref}
\usepackage{cleveref}
\usepackage[utf8]{inputenc}
\setcitestyle{numbers,square}
\usepackage[caption=false]{subfig}
\usepackage{orcidlink}

\def\dd{\mathrm{d}}

\begin{document}

\title{On the validity of the continuity equation in a modified gravity framework with CMB, DES $3\times$2pt and tomographic ISW data}
\author{Ziad Sakr$^{*,1,2,3}$\orcidlink{0000-0002-4823-3757}}
\author{Miguel Quartin$^{4,5,6,7}$\orcidlink{0000-0001-5853-6164}}

\email{$^*$ziad.sakr@net.usj.edu.lb}

% \author[a,b,c]{Ziad Sakr,}
% \author[a,d,f]{and Miguel Quartin}
\affiliation{$^1$ Instituto de Física Teórica UAM-CSIC, Campus de Cantoblanco, 28049 Madrid, Spain}
\affiliation{$^2$ Institute of Theoretical Physics, Philosophenweg 16, Heidelberg University, 69120, Heidelberg, Germany}
\affiliation{$^3$ Faculty of Sciences, Universit\'e St Joseph; Beirut, Lebanon}
\affiliation{$^4$ Centro Brasileiro de Pesquisas Físicas, 22290-180, Rio de Janeiro, RJ, Brazil}
\affiliation{$^5$ Instituto de Física, Universidade Federal do Rio de Janeiro, 21941-972, Rio de Janeiro, RJ, Brazil}
\affiliation{$^6$ Observatório do Valongo, Universidade Federal do Rio de Janeiro, 20080-090, Rio de Janeiro, RJ, Brazil}
\affiliation{$^7$ PPGCosmo, Universidade Federal do Espírito Santo, 29075-910, Vitória, ES, Brazil}

\begin{abstract}
    In this work we propose a phenomenological modification to the continuity equation at the linear perturbation level and test it using combinations of the CMB temperature, polarization and lensing potential angular spectrum, the ISW-galaxy cross power spectrum and the 3$\times$2pt lensing and galaxy clustering from DES survey. We investigate two parametrisations of this modification, both proportional to a new parameter $A_c$, which is assumed to be either constant in time, or proportional to the scale factor $a$, in order to be more relevant at late times. We find DES and ISW data to be consistent with the standard continuity equation when $A_c$ is constant, but 2--3$\sigma$ hints of a non-zero modification appear when Planck data is included. The model $A_c \propto a$ results in stronger tensions. We also test the effects of including the common extra parameters $\mu$ and $\eta$ that modify the Poisson equation and Weyl potential. Although  $A_c$, $\mu$ and $\eta$ are correlated, we still find non-zero $A_c$ when Planck data is included or without Planck if $A_c \propto a$ and only either $\eta$ or $\mu$ are allowed to vary. We conclude that violations of the continuity equation should be considered with care when testing additional deviations from general relativity.
\end{abstract}

%\pacs{03.75.Nt, 11.27.+d, 95.35.+d}
%\maketitle

\section{Introduction}\label{sec:Intro}

Despite the success of $\Lambda$CDM and General Relativity (GR) \cite{Planck:2018vyg,DES:2021wwk}, in particular when considering a cosmological constant to describe the observed accelerated expansion of our Universe, and regardless of the possible presence of tensions at the background evolution level, with the discrepancy found in the local value of the Hubble constant when compared to that inferred from other probes \cite{Riess:2021jrx}, or the recent preference for a dynamical dark energy found when CMB, Supernovae and BAO probes are combined \cite{DESI:2024mwx}, there is also still motivation to investigate effects of modifications to GR on the process of formation of large scale structures (LSS), since the latter are mainly governed by the growth of primordial perturbations till nowadays. In that regard, there is still room within the current large scale structure growth measurements for models that predicts deviation of the growth from its $\Lambda$CDM fiducial value \cite{Sakr:2023xnw}. Moreover, there has been lately found another tension, from the discrepant values on the matter fluctuation related parameter $S_8$ when constrained from weak lensing (WL) shear correlation measurements in comparison to that inferred from cosmic microwave background (CMB) \cite{KiDS:2020ghu}, or more directly the $\sigma_8$-$\Omega_{\rm m}$ confidence regions when the latter constraints are compared with those obtained from cluster counts \cite{Planck:2015lwi,Sakr:2023hrl}. We also note the possible discrepancy on the integrated Sachs–Wolfe (ISW) signal measured by stacking CMB temperature maps on positions of large voids and clusters and found in tension with predictions from $\Lambda$CDM. Several stacking analyses report a much larger temperature decrement (for voids) and increment (for clusters) than expected, both in amplitude and in statistical significance (\citep{Granett:2008ju,Szapudi:2014zha}; while other studies using different catalogs or mock tests find results consistent with standard ISW expectations or attribute the excess to selection effects and analysis choices \cite{Nadathur:2016hky}.

Often at linear to mildly non-linear scales ($ k \sim \mathcal{O}(0.001)- \mathcal{O}(0.3)\, h \,\rm Mpc^{-1}$), to which most of the current observations are still limited, the evolution of density perturbations within a theory of gravity can be described by the following quantities: the Newtonian potential $\Psi(k,z)$, the Weyl potential $\Phi(k,z)$, the density field $\rho(k,z)$, and the velocity field $v(k,z)$. These quantities are not fully independent since the first two are supposed equal within GR with deviation parametrised through the anisotropic stress parameter $\eta$, while the Newtonian potential and density contrast field are connected via the Poisson equation, with deviation from GR parametrised through the parameter $\mu$. These equations are complemented by one linking the potential and velocity field through the Euler equation, while the time derivative of the density contrast field is connected to the velocity field through the continuity equation. 

Often modifications to general relativity are performed through the $\mu$ and $\eta$ parameters while assuming GR theory for the Euler and continuity equation. However, there have been attempts to test modifications to the Euler equation by~\cite{Castello:2022uuu}, who did so by relaxing the validity of the weak equivalence principle, thus allowing dark matter to behave differently than standard matter. They showed that redshift-space distortions commonly used are not sufficient to constrain modifications to the Poisson equation, and found that such degeneracies can be broken by measuring gravitational redshift from the galaxy distribution. Although the continuity equation has not been subject to similar investigations, there have been many works that investigated the phenomenology of the beyond Horndeski class of scalar-tensor theories of gravity through a non-vanishing $\alpha_h$ parameter. This leads, among other effects, to a coupling between the sound speeds of dark energy and matter in the Jordan frame, which modifies the continuity equation conservation in the Einstein frame. These latter studies have been conducted using either galaxy and galaxy clusters profile of their potential~\cite{Pizzuti:2021brr,Haridasu:2021hzq} or also using cosmological probes such as CMB and RSD~\cite{Traykova:2019oyx}. Nevertheless, in all these attempts, no consideration of the impact of the non conservation of the continuity equation alone was investigated. 

In this work we aim at directly testing the impact on the growth of LSS of a non conservation of the continuity equation described by a non-vanishing parameter $A_{\rm c}$, and that along with the interplay with the two other parameters commonly used to model deviation from GR, $\eta$ and $\mu$, where the latter affects the growth of structure especially on sub-horizon scales, while the former impacts more the lensing or the ISW effect from the LSS on the travelling photons, or to a lesser degree, the perturbations at super horizon scales in probes such as CMB angular temperature and polarization spectra. For that we are going to combine appropriate datasets that are expected to catch effects such as the ISW signal from CMB cross correlation with galaxy distributions or the 3$\times$2-pt photometric lensing and galaxy auto- and cross-correlation probes. Note that this signal is different from the ISW from stacking CMB temperature maps mentioned above that we do include in this work. We also combine these probes with the CMB temperature, polarization and lensing angular power spectra. 

The structure of the paper is the following: in Sec.~\ref{ER} we lay down the modelling we followed for the break of the continuity equation and its impact on our used observables. In Sec.~\ref{sec:GWL} we describe the method and datasets we use to test our theory. Then in Sec.~\ref{sec:GWL} we show the result of our Bayesian study for different cases and combinations of datasets before we discuss and conclude in Sec.~\ref{conclusion}.

\bigskip
\section{Models, methods, and datasets}\label{ER}

\subsection{Theory and models}

Under general assumptions, in the quasi-static limit, the perturbed matter stress-energy tensor equations~\cite{Pace:2019vrs} in natural units $G$ (gravitational \, constant) $ = c \, (\rm{light \, speed}) = 1$ read:
\begin{align}
    & \dot{\delta}_{\rm m} + 3H\left(c_{{\rm s},{\rm m}}^{2}-w_i\right)\delta_{\rm m}  = \nonumber\\ 
    & \;\quad  -\left[1+w_{\rm m}+\left(1+c_{{\rm s},{\rm m}}^{2}\right)\delta_{\rm m}\right] 
    \vec{\nabla}\cdot\vec{u}_{\rm m}\,-\vec{u}_{\rm m}\cdot\vec{\nabla}\delta_{\rm m}\,, \label{equ:cont-pert}\\
    & \dot{\vec{u}}_{\rm m}+2H\vec{u}_{\rm m}+(\vec{u}_{\rm m}\cdot\vec{\nabla})\vec{u}_{\rm m}+\frac{\vec{\nabla}\phi}{a^2} + \nonumber\\
    & \qquad \frac{c_{{\rm s} , {\rm m}}^{2}\vec{\nabla}\delta_{\rm m}}
    {a^2\left[1+w_{\rm m}+(1+c_{{\rm s} , {\rm m}}^{2}) \delta_{\rm m}\right]} = 0\,, \label{equ:euler-pert}\\
    & \nabla^{2}\phi = 4\pi a^2{\rho}_{{\rm m}}\left(1+3c_{{\rm s}, {\rm m}}^{2}\right)\delta_{\rm m}\,, \label{equ:poisson-pert}
\end{align}
where $H$ is the Hubble parameter and $\vec{u_{\rm m}}(\vec{x},t)$ is the comoving peculiar velocity as a function of the comoving coordinates $\vec{x}=\vec{r}/a$, the pressure perturbations to density perturbations are related through the effective sound speed $c_{{\rm s},{\rm m}}^2=\delta p_{\rm m}/\delta\rho_{\rm m}$ and where we specify a relation between pressure and density by introducing the background equation-of-state parameter $w_{\rm m}={p}_{\rm m}/{\rho}_{\rm m}$. If we want to consider the standard cold dark matter paradigm, where $w_{\rm m}$ and $c_{\rm s}$ vanish and baryons are sub-dominant at our observables scales, and if we further consider $\delta_{\rm m} \ll 1$ along with assuming a simple top-hat profile for the density perturbations, i.e. $\vec{\nabla}\delta_{\rm m}=0\,,$ we see that Eq.~\eqref{equ:cont-pert}, if we isolate to the right-hand side the derivative of $\delta$ term and multiply all the rest to the left by our factor, will be the same as Eq.~\ref{modcont}. 

%Eq.~\eqref{equ:euler-pert} and Eq.~\eqref{equ:poisson-pert} in our assumptions reduce to their form within, the equivalence principle hypothesis for the former, and $\Lambda$CDM with a possible modified gravity extension, as we allow in the next section, for the latter.

Under these approximations, we recover the standard conservation equation 
\begin{equation}
    \delta_{\rm m}'=-\frac{ik_{i}{u_{\rm m}^{i}}}{aH}\,,
\end{equation}
where a prime denotes $\dd/\dd\tau$, i.e. a derivative with respect to the conformal time $\tau$. If one goes beyond the linear regime, standard symmetry considerations made in effective field theory of the large scale structure allow for counterterms which modify the continuity equation (see e.g.~\cite{Mercolli:2013bsa}) 
\begin{equation}
    \dot{\delta}_{\rm m} +\left[1+\delta_{\rm m}\right] \vec{\nabla}\cdot\vec{u}_{\rm m}\,= - \mathcal{X}_1\frac{ \nabla^{2}\delta}{\mathcal{H}} + \mathcal{X}_2 \frac{ \nabla^{2}\theta}{\mathcal{H}^2} +\cdots \,, 
\end{equation}
where $\mathcal{X}_{1,2}$ are dimensionless generalizing parameters. However, these are next-to-leading order corrections.

In this work, instead, we investigate the effects of a simpler modification of the continuity equation which could be relevant already at linear scales. If the linear conservation equation is violated, then one can write
\begin{equation}
    \delta_{\rm m}'=-\frac{{ik_{i}{u_{\rm m}^{i}}}}{aH}+Q{\delta\phi_c}'\,,
\end{equation}
where $\delta\phi_c$ is (the perturbed part of) another field whose energy is not carried by the observed particles and $Q$ is a coupling constant. It is likely that in a linear approximation we have $\delta\phi_c=-A_{\rm c}\delta_{\rm m}$. This occurs for instance in coupled dark energy, although in that case $A_{\rm c}$ would be $k$-dependent and important only at super-sound-horizon scales, as we will see later. In this case, we arrive at the equation which summarizes our proposed modification: 
\begin{equation}\label{modcont}
    (1+A_{\rm c})\delta'=-\frac{ik_{i}{u_{\rm m}^{i}}}{aH}\,,
\end{equation}
where we absorbed $Q$ into $A_{\rm c}$. 

Our choice is also compatible with another acceptable assumption that no coupling between dark energy and dark matter exists where in that case, as already mentioned above, $A_{\rm c}$ will be $k$-dependent~\cite{Gomez-Valent:2022bku}, since one has, with $Q'$ being the CDE coupling constant $T^{\mu\nu}_{(m);\nu} = Q'T_{(m)}\phi_c^{\mu},$ and $T^{\mu\nu}_{(\phi_c);\nu} = -Q'T_{(m)}\phi_c^{\mu}$ where $m$ and $\phi_c$ stands respectively for matter and the dark energy related scalar field, so that, without loss of generality in the quasi-static limit, 
$$\delta\phi_c=-3\frac{{ aH}^2}{k^2}Q'\Omega_{\rm m}\delta$$ 
yielding 
$$A_{\rm c}=-3\frac{{ aH}^2}{k^2}{Q'}^2(1-3c_{\rm s}^2)\Omega_{\rm m}\,.$$ 
However, even in the previous case, we see that the effect is not expected to be important at horizon scales, while here we will be using probes more effectively at sub-horizon scales.

To proceed with exploring the effect of our modelling, we start by considering that the gravitational system is described by a perturbed flat FLRW line element in the Newtonian gauge
\begin{equation}
    \dd s^2 = -\left(1+2\Psi\right)\dd t^2 +a^2\left(1-2\Phi\right)\left(\dd r^2 + r^2 \dd\Omega^2\right),
\end{equation}
where $r$ is the comoving radial distance, and we have introduced the  scalar perturbations $\Psi$ and $\Phi$. The former is sourced by the mass distribution through the Poisson equation. The latter can be conveniently absorbed into the Weyl potential
\begin{equation}
    \Phi_W = \frac{\Phi+\Psi}{2}\,,
\end{equation}
which is the degree of freedom governing both gravitational lensing and ISW effects. The gravitational potentials, if we assume at first that they are slowly varying, can be mapped out through the equation of linear matter perturbations~\cite{Euclid:2025tpw}
\begin{align}
    -k^2\,\Psi(a,{k}) &= 4\pi\,a^2\,\rho_{\rm m}(a)\,\delta_{\rm m}(a,{k})\,\mu(a,k), \label{eq:mu}\\ 
    \Phi(a,{k}) &= \Psi(a,{k})\,\eta(a,k).  
    \label{eq:eta}
\end{align}
Let's further assume the validity of the equivalence principle, and both the Euler and continuity equations at deep sub-horizon scales ($ k/(aH)\gg 1$). The perturbed variables and our parameter $A_{\rm c}$, considered in a first case as constant, and imposing $\theta=-ik_{i}u_{\rm m}^{i}/aH$, would be given by:
\begin{align}
    \theta_{\rm m} & = (1+A_{\rm c})\,\delta_{\rm m}' \,, \label{eq:tetadelta}\\
    \theta_{\rm m}' & =-\left(2+\frac{H'}{H}\right)\theta_{\rm m}+\left({\frac{k}{aH}}\right)^{2}\Psi \,.
\end{align}
Combining these two equations with 
\begin{equation}
    \label{parametrization-Poisson}
    k^2\Psi=-4\pi{a^2}\mu(a,k) \rho_{\rm m} \delta_{\rm m} \,, 
\end{equation}
\begin{equation}
    \label{parametrization-Weyl}
    k^2\Phi_W= -4\pi{a^2}\mu(a,k)\big(1+\eta(a,k) \big) \rho_{\rm m} \delta_{\rm m} \,, 
\end{equation}
we obtain the modified second order growth equation:\footnote{Note that for $A_c=-1$ we get a divergent $\delta_{\rm m}'$ for non-zero $\theta_{\rm m}$. We therefore will always limit in what follows the range of allowed $A_c$ to be above this threshold.}
\begin{equation}
    \delta_{\rm m}''+\delta_{\rm m}'\left(2+\frac{H'}{H}\right)=-{\frac{\mu}{1+A_{\rm c}}}\,\rho_{\rm m}\delta_{\rm m}\,.\label{eq:Actegrowth}
\end{equation}
In this case $\mu$ and $A_{\rm c}$ posterior probability distributions should be strongly correlated if we constrain them through probes that are affected by the growth such as the ones we use in this work, to wit the weak lensing shear and the ISW galaxy correlations (see Sec.~\ref{sec:GWL}). The interplay of $A_{\rm c}$ and $\eta$ comes, in this case, from the effect that the latter has with $\mu$ on the Weyl potential present in the aforementioned probes, where $\mu$ and $\eta$ are anticorrelated. On the other hand, $A_{\rm c}$ and $\eta$ interact in a less straightforward way through their interplay in the Weyl potential and growth of structure product, which enters in our used probes. 

As discussed above, we will consider two parametrizations for the modification of the continuity equation: 
\begin{align}
    A_c^{\rm eff}(a) &= A_c = \rm{const.} \,,\\
    A_c^{\rm eff}(a) &= A_c a \,.
\end{align}
In the case where $A_{\rm c}$ is considered as a  dynamical variable, Eq.~\eqref{modcont} becomes (we will drop the ``eff'' superscript for simplicity in the equations below) 
\begin{equation}
    (1+A_{\rm c})\delta_{\rm m}'+A_{\rm c}'\delta_{\rm m}=\theta_{\rm m}\,,\\
\end{equation}
so that 
\begin{equation}
    (1+A_{\rm c})\delta_{\rm m}'' + 2 A_{\rm c}' \delta_{\rm m}' + A_{\rm c}''\delta_{\rm m} =\theta_{\rm m}' \,.
\end{equation}
The second order growth equation becomes now:
\begin{equation}
    \begin{aligned}
    &(1+A_{\rm c})\delta_{\rm m}''+\bigg[2 A_{\rm c}'+(1+A_{\rm c})\left(2+\frac{H'}{H}\right)\bigg]\delta_{\rm m}'= \\ 
    &\qquad \bigg[-{\mu} -A_{\rm c}'' -A_{\rm c}' \left(2+\frac{H'}{H}\right)\bigg] \rho_{\rm m} \delta_{\rm m} \,.\label{eq:Avargrowth}
    \end{aligned}
\end{equation}\\
In this case $A_{\rm c}$ is not necessarily correlated with $\mu$ (unless its derivative is small with respect to it), while the correlation between $\mu$ and $\eta$ remains the same because they are both included in the modification of the Weyl potential. However, now that $A_{\rm c}$ is time-dependent, we could expect it to be anticorrelated with $\eta$ through the relation between $\eta$ and the derivative of the growth in the ISW formalism.

In this work however, we may additionally combine in some cases with the CMB angular power spectrum on scales larger than the horizon, and the effect of $A_{\rm c}$, $\mu$ and $\eta$, parameterised as constant or varying with redshift, would be implemented through \texttt{MGCLASS} in which case the quasi-static approximation is not relaxed as in the above equations, even though it has been shown that the super-horizon treatment of the effect of $\mu$ and $\eta$ could be separated from that on the sub-horizon in the equations implemented in \texttt{MGCLASS}.

The analytic treatment when $A_{\rm c}$, $\mu$ and $\eta$ are taken as dynamical is too complicated to give direct insights on the interplay between these parameters so we only present here the constant parameterisation case, where without the quasi-static approximation, an additional component enters Eq.~\eqref{eq:tetadelta}, which becomes 
\begin{equation}
    \theta_{\rm m}  = (1+A_{\rm c})\,\delta_{\rm m}' + 3 \Phi ' \,.
\end{equation}
Eq.~\eqref{eq:mu} has to be written now as
\begin{equation}
    \label{parametrization-Poisson-gauge}
    k^2\Psi=-{a^2}\mu(a,k) \rho_{\rm m} \Delta_{\rm m} \,, 
\end{equation}
where $\Delta_{\rm m} = \delta + 3 (a H/k) \theta$. Defining $\,u = (k/(aH) ) \theta\,$ and $\,E_z = H^2/H_0^2$, we arrive at the following system of equations which generalize those of e.g.~\cite{Pogosian:2010tj} for $A_c \neq 0$ 
\begin{align}
    \label{eq:Deltapr}
    \Delta_{\rm m}'&= \frac{-\frac{9}{2}\frac{\Omega_{\rm m}}{a^3 E_z} \mu\left[{1 -\frac{\eta}{1+A_{\rm c}}} \right] \Delta_{\rm m}
    +\left[ 3\frac{H'}{H} - \frac{\left(\frac{k}{aH} \right)^2}{1+A_{\rm c}} \right] u
    }{{\left(\frac{k}{aH} \right)}^2 +\frac{9}{2} \frac{\Omega_{\rm m}}{a^3 E_z} \frac{\eta \mu}{1+A_{\rm c}}} \,,\\
    u'&=-\left[2+\frac{H'}{H} \right] u -\frac{3}{2}\frac{\Omega_{\rm m}}{a^3 E_z} \mu \Delta_{\rm m}\,.
    \label{eq:vpr}
\end{align}
Dividing our treatment between sub- and super-horizon scales, we see that for the sub-horizon scales Eq.~\eqref{eq:Deltapr} yields
\begin{equation}
    \Delta_{\rm m}'= - \frac{u}{1+A_{\rm c}}\,,
\end{equation}
so that we end up with an equation similar to \eqref{eq:Actegrowth}
\begin{equation}
    \Delta_{\rm m}''+\Delta_{\rm m}'\bigg(2+\frac{H'}{H}\bigg)+{\frac{\mu}{1+A_{\rm c}}}\,\rho_{\rm m}\Delta_{\rm m} = 0\, . \label{eq:ActeDeltagrowth}
\end{equation}
Again $\mu$ and $A_{\rm c}$ are expected to be correlated, while for the super-horizon scales, Eq.~\eqref{eq:Deltapr} becomes 
\begin{eqnarray}
    \label{SH-conservation}
    \Delta_{\rm m}'&=& -\left[{{1+A_{\rm c} \over \eta} -1} \right]\Delta_{\rm m}+{2 \over 3}{H'H (1+A_{\rm c})\over \eta \mu E_{\rm m}}u  \,,
\end{eqnarray}
where $E_{\rm m}= \Omega_{\rm m} H_0 / a^3$. This, combined with Eq.~\eqref{eq:vpr}, yields
\begin{align}
    \Delta_{\rm m}''&{C \eta \mu \over (1+A_{\rm c})}
    +{\eta \mu \over (1+A_{\rm c})} \left[-C' + C\left(\frac{1+A_c}{\eta}-1\right)  \right] \Delta_{\rm m}' \nonumber \\    
    &+\left({{1+A_{\rm c} \over \eta} -1} \right){C' \eta \mu \over (1+A_{\rm c})}\Delta_{\rm m}+\left(2+\frac{H'}{H}\right) \times \nonumber\\
    &\left( {C (1+A_{\rm c}) \over \eta \mu}\Delta_{\rm m}'
    +\left({{1+A_{\rm c} \over \eta} -1} \right){C (1+A_{\rm c}) \over \eta \mu}\Delta_{\rm m}\right)\nonumber \\
    & +{3 \over 2}{E_{\rm m} \over H^2} \mu \Delta_{\rm m} = 0\,,\label{eq:ActeSHDeltagrowth}
\end{align}
where $C = (3 E_{\rm m})/(2 H'H)$ was defined to simplify the presentation of the derivative of $u$ to keep the relevant terms clear. Here we see that $\eta$ and $A_{\rm c}$ should be correlated, though the effect might be cancelled from their interplay in our probes. On the other hand $\mu$ and $\eta$ should be anti-correlated,  and no specific relation is clear between $\mu$ and $A_{\rm c}$.

\bigskip

\section{Methods and datasets}\label{sec:methanddata}

\noindent As part of the baseline we use in this work, we include the cross correlation of ISW and CMB, the formulation of which is given by~\cite{Douspis:2008xv,Krolewski:2021znk}
\begin{equation}
    C_{\ell}^{(Tg)}=\frac{2}{\pi}\int k^{2}dkP(k)K_{\ell}^{\Phi}K_{\ell}^{g} \,,
\end{equation}
with a FLRW $\Lambda$CDM background assumed we have
\begin{align}
    K_{\ell}^{g} & =\int \dd z\frac{\dd N}{\dd z}b(z) D(z)j_{\ell}(k\chi(z)) \,, \\
    K_{\ell}^{\Phi} & \!=\!\frac{3\Omega_{\rm m}H_{0}^{2}}{k^{2}} \!\int \!\dd z\frac{d}{dz}\left [\frac{\mu}{2}(1+\eta)(1+z)D(z)\right ]j_{\ell}(k\chi(z)) ,
\end{align}
where $b(z)$ is the bias, $D(z)$ the growth factor, $j_{\ell}$
the spherical Bessel functions, $\chi(z)$ the comoving distance, $\dd N(z)/\dd z$ the normalized galaxy redshift distribution, and $\eta$ parametrizes modification of gravity. For $\eta=\mu=1$ we recover the standard GR equation. Following~\cite{Stolzner:2017ged}, our likelihood for $C_{\ell}^{(Tg)}$ includes a free parameter $A_{\rm ISW}$ that quantifies the agreement between the data and the $\Lambda$CDM model and has unity value for $\Lambda$CDM.

We also combine constraints with those from the 3$\times$2pt power spectrum of the galaxy lensing, clustering and their cross correlations, in which the observed angular lensing-lensing convergence power spectrum from divided into several redshift bins can be written as 
\begin{equation}
    C_{ij}^{\rm \gamma\gamma}(\ell)= \int{\rm d}z\,\frac{W_i^{\rm \gamma}(z)W_j^{\rm \gamma}(z)}{H(z)r^2(z)}\frac{\mu}{2}(1+\eta)P_{\rm m}^{\rm }(k,z)\,\label{eq:Cl_GCph}a\,,
\end{equation}
where $P_{\rm m}(k,z)$ is the galaxy-galaxy power spectrum, evaluated at $k=k_{\ell}(z)=\frac{\ell + 1/2}{r(z)}$, and
\begin{align}
    W_i^{\rm \gamma}(k,z) \!=\! \frac{3}{2}\Omega_{\rm m} {H_0^2}(1+z)r(z) \! 
    \int_z^{z_{\rm max}}{\!\!\!\! \dd z'\frac{n_i(z)}{\bar{n}_i}\frac{r(z')\!-\!r(z)}{r(z')}} . \label{eq:wlmg}
\end{align}
Above, $n_i(z)/\bar{n}_i$ and $b_i(k,z)$ are, respectively, the normalised galaxy distribution and the galaxy bias in the $i$-th redshift bin. We also include intrinsic alignment effects, where the weight function can be written as
\begin{equation}
    W_i^{\rm IA}(z) =  \frac{n_{i}(z)}{1/H(z)} = {H_0} n_i(z) E(z){\cal{A_{\rm c}}}_{\rm IA} {\cal{C}}_{\rm IA} \Omega_{\rm m}\frac{\mathcal{F}_{\rm IA}(z)}{G(z,k)} \,,
    \label{eq:iaweight}
\end{equation}
where  
$${\cal{F}}_{\rm IA}(z) = (1 + z)^{\eta_{\rm IA}}[\langle L \rangle(z)/L_{\star} (z)]^{\beta_{\rm IA}}, $$
to obtain, respectively, the correlation between background shear and foreground intrinsic alignment $C^{\rm I\gamma}_{ij}(\ell)$, and the autocorrelation of the foreground intrinsic alignment $C^{\rm II}_{ij}(\ell)$. We also include the photometrically detected galaxy-galaxy correlations for which we replace in Eq.~\eqref{eq:Cl_GCph} the weights by the radial function for galaxy clustering defined as
\begin{equation}\label{eq:photoGCwin}
    W^{\rm G}_i (k,z)=b_i(k,z) n_i (z) H(z)\, .
\end{equation}
Finally, we investigate constraints obtained from combinations with the CMB temperature, polarization, and their cross correlations C$_\ell$ and lensing spectrum D$_\ell$ probe from the data released by the Planck satellite mission \cite{Planck:2018vyg} (hereafter Plk18).

We run our MCMC chains using \texttt{MGCLASS} II \cite{Sakr:2021ylx} which is interfaced with the cosmological data analysis code \texttt{MontePython} \cite{Brinckmann:2018cvx} in which we use the publicly available likelihood \cite{Planck:2019nip} that contains the Planck high-$\ell$ (for high-temperature), Planck low-$\ell$ and low-EE  (for low-temperature and polarization), and Planck lensing (CMB lensing reconstruction). We also make use within the inference code of the CMB$\times$ISW likelihood from \cite{Stolzner:2017ged} and adapted it to incorporate our model testing. As tracers of matter distribution we use five catalogues of extragalactic sources: the 2MASS Photometric Redshift catalogue 2MPZ \citep{Bilicki:2013sza}, the WISE × SuperCOSMOS photo-z catalogue (henceforth WI$\times$SC)~\citep{Bilicki:2016irk}, the SDSS DR12 photometric dataset \citep{2016MNRAS.460.1371B}, a catalogue of photometric quasars compiled from the SDSS DR6 dataset \citep{Richards:2008eq} and the NRAO VLA Sky radio sources Survey NVSS \citep{Condon:1998iy}. Finally, for the 3$\times$2pt lensing and clustering correlations, we use data from the dark energy survey (DES) collaboration~\cite{DES:2017myr,DES:2021wwk} with the DES likelihood implemented as well by us based on the official public one, where we limit ourselves to the linear scales following the procedure of~\cite{Abbott:2018xao}.

Table~\ref{tab:params} summarizes all free parameters (including the nuisance ones) and their assumed priors in three groups: the cosmological and modified gravity ones, the ones used in the DES $3\times2$pt likelihood, and those used in the tomographic ISW likelihood.

\bigskip
\section{Constraints on the continuity equation and the anisotropic stress parameters}\label{sec:GWL}

In this section we show MCMC constraints on the cosmological parameters with the $A_{\rm c}$ parameter, either alone or when we consider further extensions from the standard scenario by including different combinations of $A_{\rm c}$, $\eta$ and $\mu$ as free parameters. We also allow them to be either a constant or time-varying according to a fixed dynamical scale parameterisation. We provide results for different combinations of our probes, namely the cross correlation between ISW and CMB, and the 3$\times$2pt joint lensing and clustering probe from DES galaxy survey. 

In this first scheme, we additionally consider priors on some of the cosmological parameters that will not be constrained by the sub-horizons probes, using as reference the CMB power spectrum results. In particular, we do so for these parameters: the spectral index $n_{\rm s}$ and the baryon matter density $\Omega_{\rm b}$. Subsequently, in a second bundle, we follow the same previous scheme, but we further combine results with the CMB temperature, polarization and lensing power spectrum and add different constraining observables coming from the effect of the standard cosmological model extension parameters on the late ISW and lensing effect impact on the CMB spectrum.

\begin{table}[t]
\caption{List of parameters and priors. We use either uniform (${\cal U}[a, b]$) or Gaussian (${\cal N}(\mu, \sigma)]$) priors for all parameters. The Gaussian priors were used when we only consider Plk18 priors. \emph{Top:} Cosmological and modified gravity parameters.
\emph{Center:} DES 3$\times$2pt likelihood. \emph{Bottom:} CMB $\times$ ISW likelihood.}
\begin{center}
\begin{tabular}[t]{| c  c |}
\hline
Parameter & Prior \\  
\hline 
\multicolumn{2}{|c|}{{\bf Cosmology}} \\
$\Omega_{\rm m}$  &  ${\cal U}$[0.1, 0.9]\\ 
$\ln (10^{10} A_{\rm s})$ &  ${\cal U}$[$1.6,3.9$] \\ 
$n_{\rm s}$ &  ${\cal U}$[0.87, 1.07], ${\cal N}$(0.965, 0.004) \\
$\Omega_{\rm b}$ &  ${\cal U}$[0.03, 0.07], ${\cal N}$(0.0493, 0.0001)\\
$h$  &  ${\cal U}$[0.3, 0.91]  \\
$\mu$  & ${\cal U}$[-1.0, 1.0] \\
$\eta$  & ${\cal U}$[-1.0, 1.0] \\
$A_{\rm c}$ &   ${\cal U}$[-1.0, 1.0]   \\
\hline
\end{tabular}
\smallskip\\
\begin{tabular}[t]{| c  c |}
\hline
Parameter & Prior \\  
\hline 
\multicolumn{2}{|c|}{{\bf Lens Galaxy Bias}} \\
$b_{i} (i=1,5)$   & ${\cal U}$[0.8, 3.0] \\
\hline
\multicolumn{2}{|c|}{{\bf Intrinsic Alignment}} \\
\multicolumn{2}{|c|}{{$A_{\rm IA}(z) = A_{\rm IA} [(1+z)/1.62]^{\eta_{\rm IA}}$}} \\
$A_{\rm IA}$   & ${\cal U}$[$-5,5$] \\
$\eta_{\rm IA}$   & ${\cal U}$[$-5,5$] \\
\hline
\multicolumn{2}{|c|}{{\bf Lens photo-$z$\ shift (red sequence)}} \\
$\Delta z^1_{\rm l}$  & ${\cal U}$[$0.008, 0.007$] \\
$\Delta z^2_{\rm l}$  & ${\cal U}$[$-0.005, 0.007$] \\
$\Delta z^3_{\rm l}$  & ${\cal U}$[$0.006, 0.006$] \\
$\Delta z^4_{\rm l}$  & ${\cal U}$[$0.000, 0.010$] \\
$\Delta z^5_{\rm l}$  & ${\cal U}$[$0.000, 0.010$] \\
\hline
\multicolumn{2}{|c|}{{\bf Source photo-$z$\ shift}} \\
$\Delta z^1_{\rm s}$  & ${\cal U}$[$-0.001, 0.016$] \\
$\Delta z^2_{\rm s}$  & ${\cal U}$[$-0.019, 0.013$] \\
$\Delta z^3_{\rm s}$  & ${\cal U}$[$+0.009, 0.011$] \\
$\Delta z^4_{\rm s}$  & ${\cal U}$[$-0.018, 0.022$] \\
\hline
\multicolumn{2}{|c|}{{\bf Shear calibration}} \\
$m^{i}_{\rm metacalibration} (i=1,4)$ & ${\cal U}$[$0.012, 0.023$]\\
$m^{i}_{\rm IM3SHAPE_{}} (i=1,4)$ & ${\cal U}$[$0.0, 0.035$] %\vspace{.02cm}
\\ 
\hline
\end{tabular}
\smallskip\\
\begin{tabular}[t]{| c  c |}
\hline
Parameter & Prior \\ 
\hline
\multicolumn{2}{|c|}{{\bf ISW calibration parameter}} \\
$A_{\rm ISW}$   & ${\cal U}$[0.85, 2.3] \\
\hline 
\multicolumn{2}{|c|}{{\bf MPZ}} \\
$b_0$   & ${\cal U}$[1.2,   1.8] \\
$b_1$   & ${\cal U}$[1.1,   1.7] \\
$b_2$   & ${\cal U}$[1.5,   2.3] \\
\hline
\multicolumn{2}{|c|}{{\bf SDSS}} \\
$b_0$   & ${\cal U}$[0.9,   1.4] \\
$b_0$   & ${\cal U}$[0.7,   1.05] \\
$b_1$   & ${\cal U}$[0.6,   0.95] \\
$b_2$   & ${\cal U}$[0.85,   1.3] \\
$b_2$   & ${\cal U}$[0.7,   1.3] \\
\hline
\multicolumn{2}{|c|}{{\bf WI$\times$SC}} \\
$b_0$   & ${\cal U}$[0.8,   1.3] \\
$b_1$   & ${\cal U}$[0.65,   1.1] \\
$b_2$   & ${\cal U}$[0.8,   1.2] \\
\hline
\multicolumn{2}{|c|}{{\bf QSO}} \\
$b_0$   & ${\cal U}$[0.6,   1.8] \\
$b_1$   & ${\cal U}$[1.0,   2.6] \\
$b_2$   & ${\cal U}$[1.7,   3.6] \\
\hline
\multicolumn{2}{|c|}{{\bf NVSS}} \\
$b_0$   & ${\cal U}$[1.9,   3.0] \\
\hline
\end{tabular}
\end{center}
\label{tab:params}
%\label{tab:DESparams}
\end{table}

\begin{figure*}[t]
\centering
\includegraphics[width=.7\textwidth]{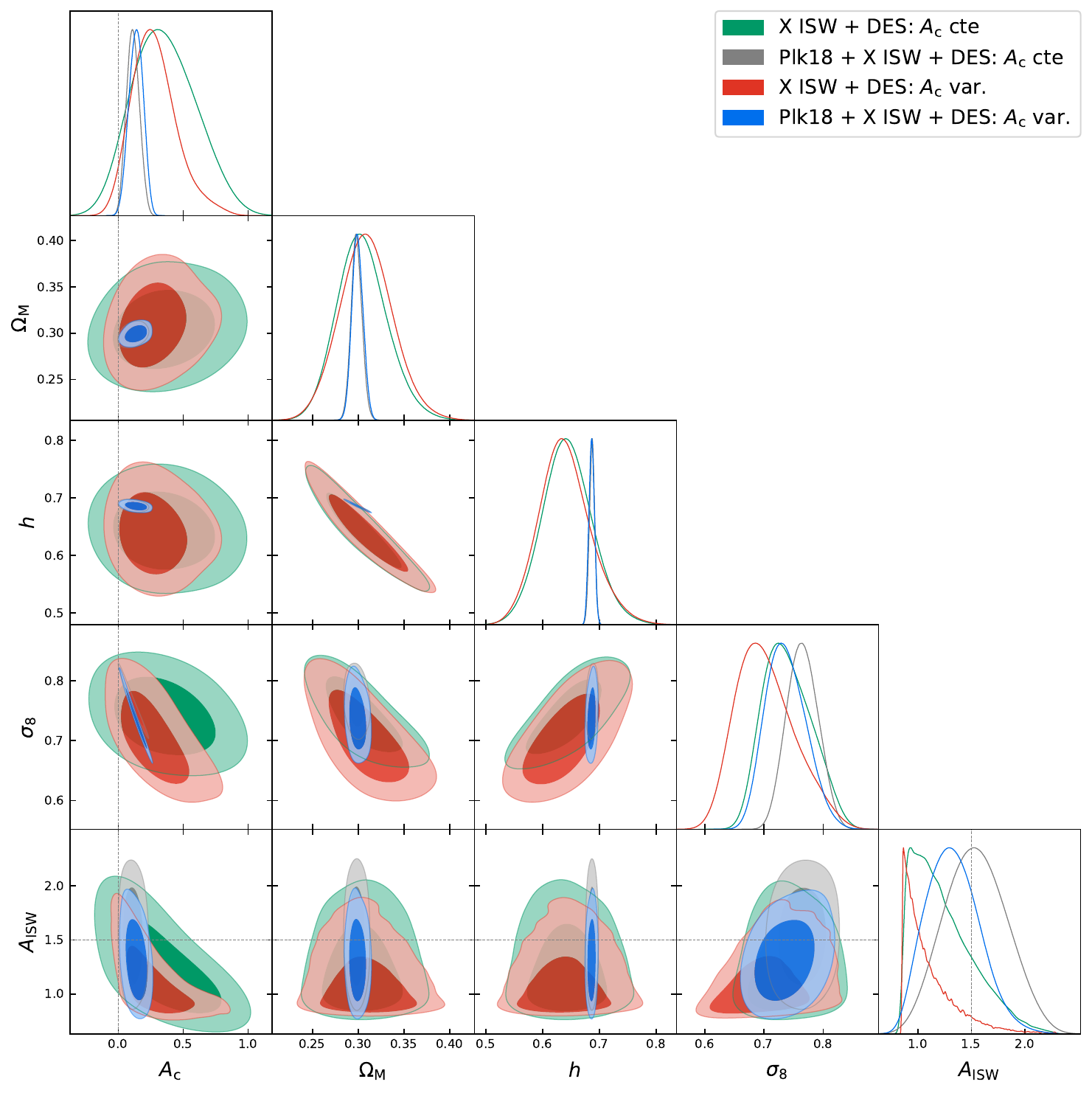}
\caption{68\% and 95\% confidence contours for $\Omega_{\rm m}$, $h$, $A_{\rm ISW}$, $\sigma_8$, and $A_{\rm c}$ inferred using ISW cross correlation with galaxy surveys distributions, and 3$\times$2pt photometric lensing and galaxy correlations, and cross correlations from DES survey, with and without CMB constraints. We note that the inclusion of CMB $TTTEEE$+lensing results in non-zero constraints for $A_c$ above the 2$\sigma$ confidence-level. The prior ranges for the different parameters follow Table~\ref{tab:params}.}
\label{fig:ANCalone}
\end{figure*}

Table~\ref{tab:params} summarizes the parameters employed here, and the priors assumed for each of them. The left sub-table show the cosmological and model extension parameters and priors. The center and right sub-tables, show the same as used by the DES 3$\times$2pt likelihood \cite{DES:2017myr,DES:2021wwk} and CMB $\times$ ISW likelihood in \cite{Stolzner:2017ged}, respectively (see Sect.~\ref{sec:methanddata} for details on the datasets used).  We considered uniform (${\cal U}[a, b]$) priors for all parameters or Gaussian (${\cal N}(\mu, \sigma)$) when we only consider Plk18 priors.

We start by showing, in Figure~\ref{fig:ANCalone}, the results of our MCMC runs when allowing $A_{\rm c}$ alone as an additional extension to the $\Lambda$CDM model, and that for two sorts of probes combination, to wit one using only ISW-GC and 3$\times$2pt WLGC with  CMB priors, and one when additionally adding the Planck likelihood itself. We observe in the first case, for either constant or dynamical $A_{\rm c}$, that its bounds are still compatible with a $\Lambda$CDM-like null value, while the constraints on $h$ and $\sigma_8$ explore a range that covers both their local and CMB values. We also notice that the $A_{\rm ISW}$ bounds are in this case compatible with the unity value, shifted from those obtained in~\cite{Stolzner:2017ged} within the $\Lambda$CDM model. 

When we combine with the full CMB likelihood, we observe a tightening of the constraints, and we re-obtain the current tension in $h$ with its local value, while $\sigma_8$ bounds are still broad. Interestingly, the $A_{\rm c}$ bounds are in this case strongly constrained and prefer a value of $\sim0.15$ at over 2$\sigma$ from the fiducial null value. This is mainly due to the discrepancy on the $A_{\rm ISW}$ value between CMB and local probes, resulting in tight constraints on $A_{\rm c}$ when combined.

\begin{figure*}[t]
\centering
\includegraphics[width=0.85\textwidth]{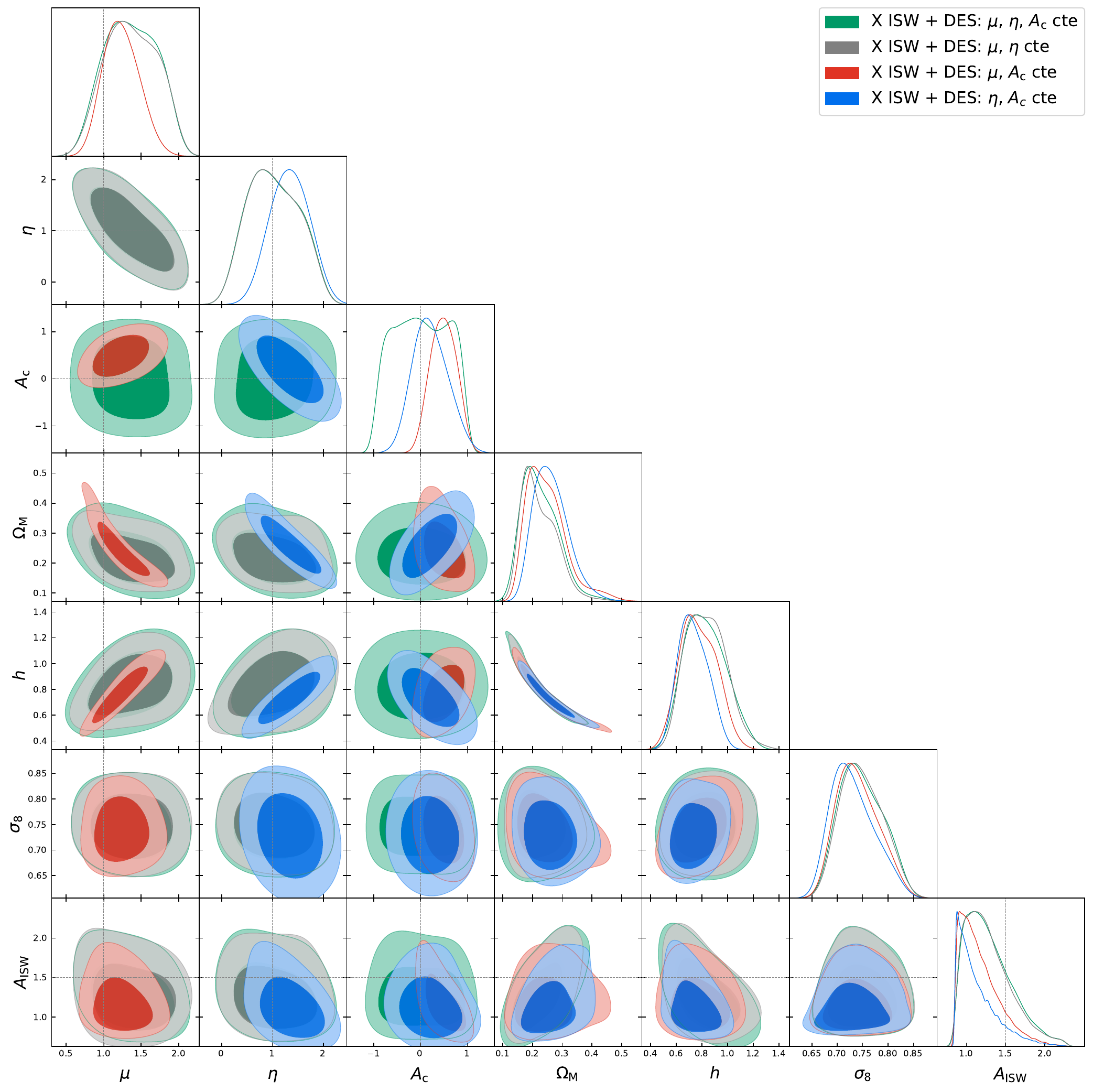}
\caption{Same as Figure~\ref{fig:ANCalone} but adding also the $\mu$ and $\eta$ modified gravity parameters and assuming $A_c$, $\mu$ and $\eta$ to be constant in time. \bigskip}
\label{fig:cteprio}
\end{figure*}

\begin{figure*}[t]
\centering
\includegraphics[width=0.85\textwidth]{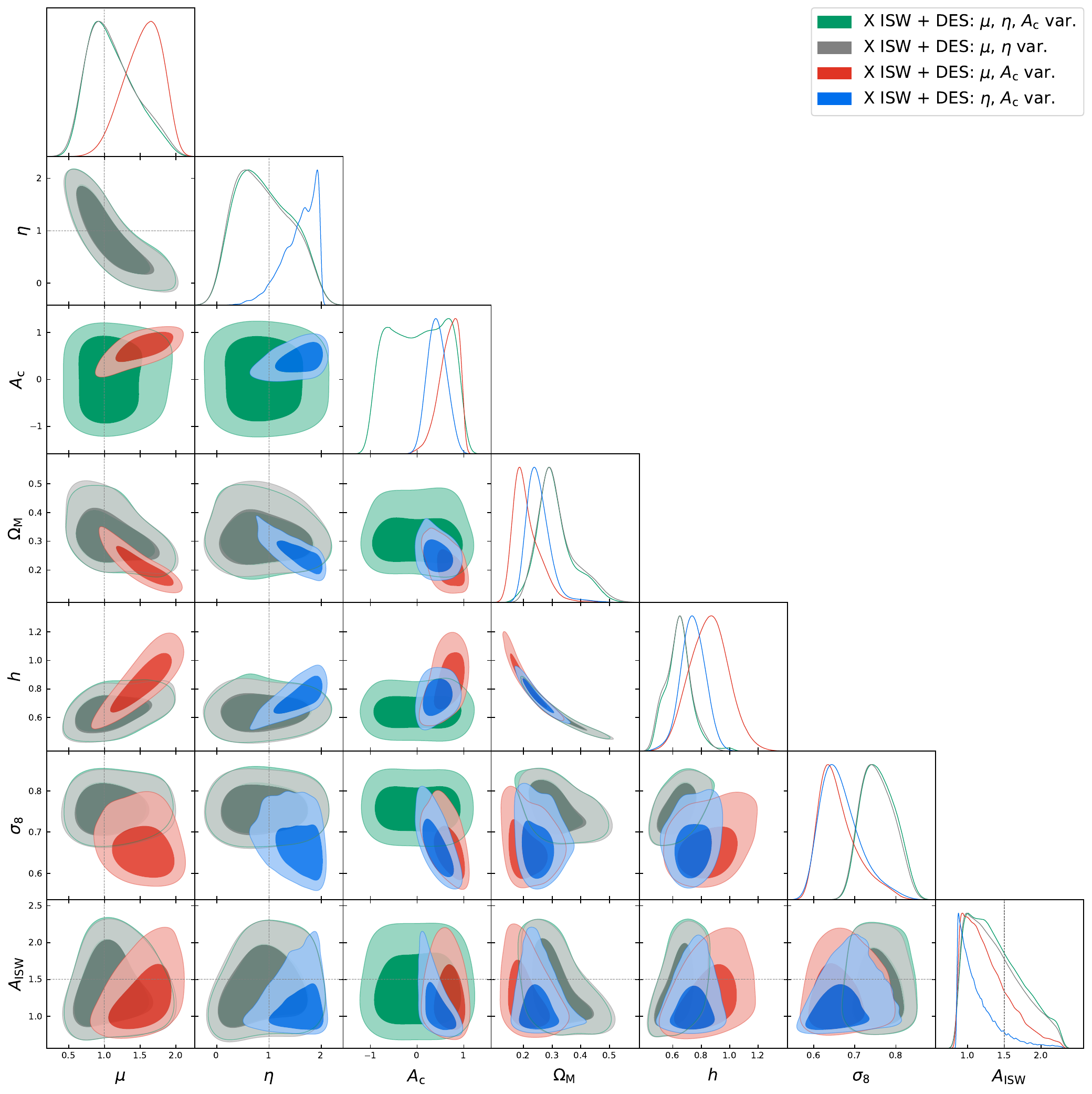}
\caption{Same as Figure~\ref{fig:cteprio} but now allowing time-variations of $A_c$, $\mu$ and $\eta$. \bigskip}
\label{fig:varprio}
\end{figure*}

Let's now move on to cases where we allow for a more general modified gravity scenario, and include the $\mu$ and/or $\eta$ parameters, in addition to $A_c$ and $A_{\rm ISW}$. Just as for $A_c$, we will consider two models, one in which $\mu$ and $\eta$ are constant, and one in which they are proportional to $a$. So we either consider all three $A_c$, $\mu$ and $\eta$ to be all constant, or all time-varying.

We start with the constant $A_c$ model. We depict in Figure~\ref{fig:cteprio} the resulting cornerplot for the case where we combine ISW-GC and 3$\times$2pt WLGC along with priors from CMB Planck 2018. We analyze four different cases, varying: (i) $A_{\rm c}$ and $\eta$ (blue lines); (ii) $A_{\rm c}$ and $\mu$ (red lines); (iii)  $\eta$ and $\mu$ but no $A_c$ (gray lines); (iv) all four together (green lines). We also show constraints on the classical $\mu$-$\eta$ combination when both are left free to vary (gray lines). We note in this case that the extra degrees of freedom result in poor constraints for $A_{\rm c}$, $\eta$ and $\mu$, and that in all cases they are consistent with the $\Lambda$CDM scenario. Moreover, as expected and explained in Sec.~\ref{sec:methanddata}, we observe for a constant $A_c$ that $A_{\rm c}$ and $\mu$ are correlated while the opposite occurs between $A_{\rm c}$ and $\eta$ due, in the former case, to their interplay in the growth equation, and in the latter one to that of how $A_{\rm c}$ and $\eta$ enter into the Weyl potential expression. 

We then test the $A_c \propto a$, $\eta \propto a$, $\mu \propto a$ model with the same dataset. Results are illustrated in Figure~\ref{fig:varprio}. The correlation trend above is modified in the dynamical $A_c$ case, and $A_{\rm c}$ and $\eta$ are now also positively correlated. This is due to their relation in the ISW equation where the differentiation of $\eta$, now dynamical, affects its evolution. Note that the correlation between $\mu$ and $\eta$ is always positive for the two parameterisations because it is mainly driven by their interplay in the Weyl potential entering both sub-horizon probes. Leaving free all our model extension parameters alter and weaken these trends where it seems that the $\eta$-$\mu$ correlation is strongly preferred in either of the adopted parameterisations such that both vary almost independently of $A_{\rm c}$. We also observe, for almost all cases and parameterisations, that the bounds on $A_{\rm c}$ show preference for positive values, with however a preference for the null value for the $A_{\rm c}$-$\eta$ case while the symmetry is fully restored when we leave all three parameters free at once. 

We also observe in both constant and time-varying models that $A_{\rm ISW}$ prefer a unity value, contrary to some results in the literature. In the constant model, we see that $\Omega_{\rm m}$ or $h$ or $\sigma_8$ bounds widen with respect to $\Lambda$CDM to the point that no tension is seen, at the cost of course of a lost in the constraining power. In the dynamical parameterisation however, $\sigma_8$ bounds are showing preference for low values for the $A_{\rm c}$-$\eta$ and $A_{\rm c}$-$\mu$ cases, in tension with those obtained with the $\mu$-$\eta$ and $A_{\rm c}$-$\eta$-$\mu$ free cases.

\begin{figure*}[t]
\centering
\includegraphics[width=0.85\textwidth]{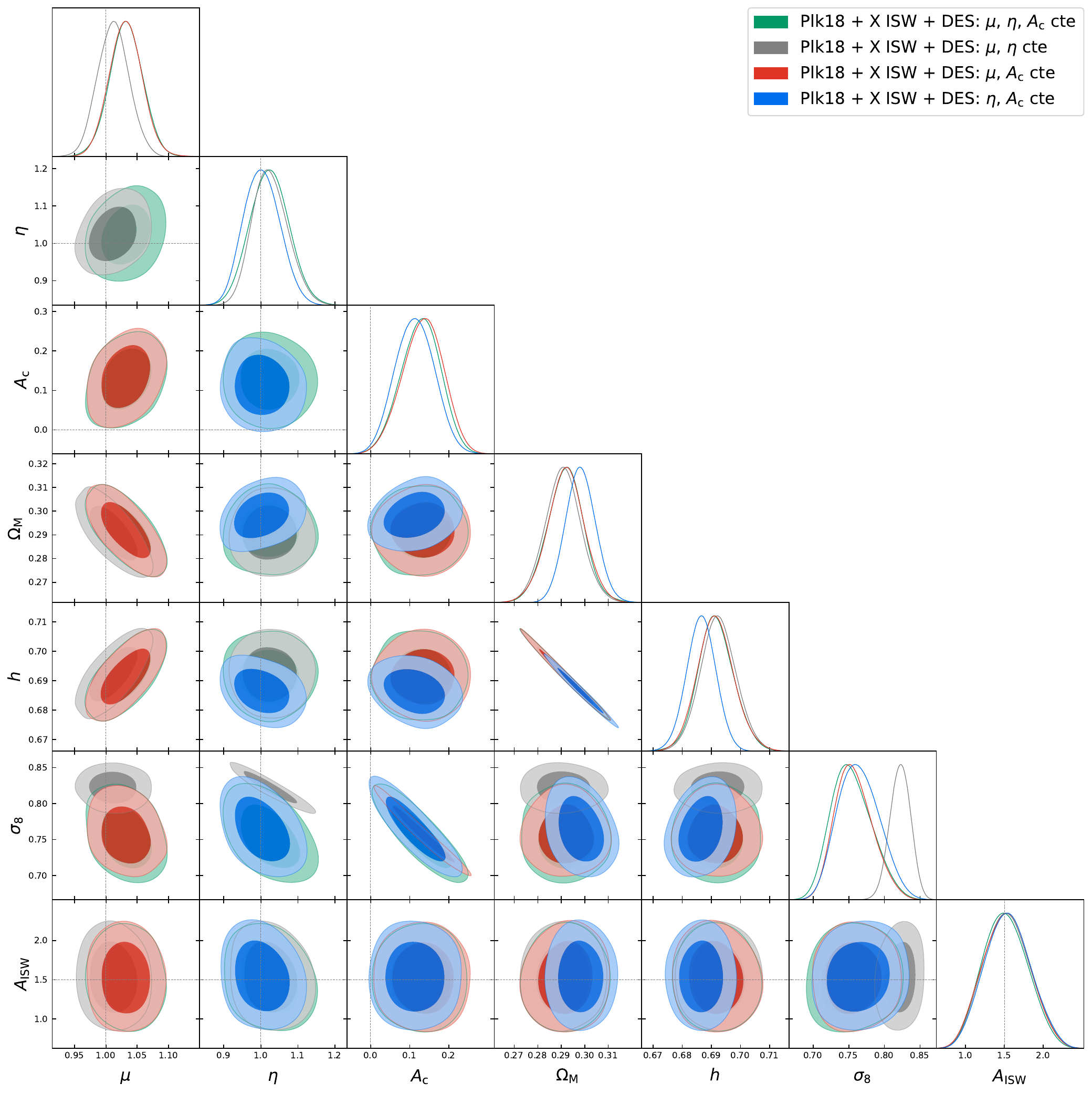}
\caption{Same as Figure~\ref{fig:cteprio} but now adding different combinations of CMB $C_{\ell}^{TT,TE,EE}$ and its lensing potential from Plk18. \bigskip}
\label{cteCMB}
\end{figure*}

\begin{figure*}[t]
\centering
\includegraphics[width=0.85\textwidth]{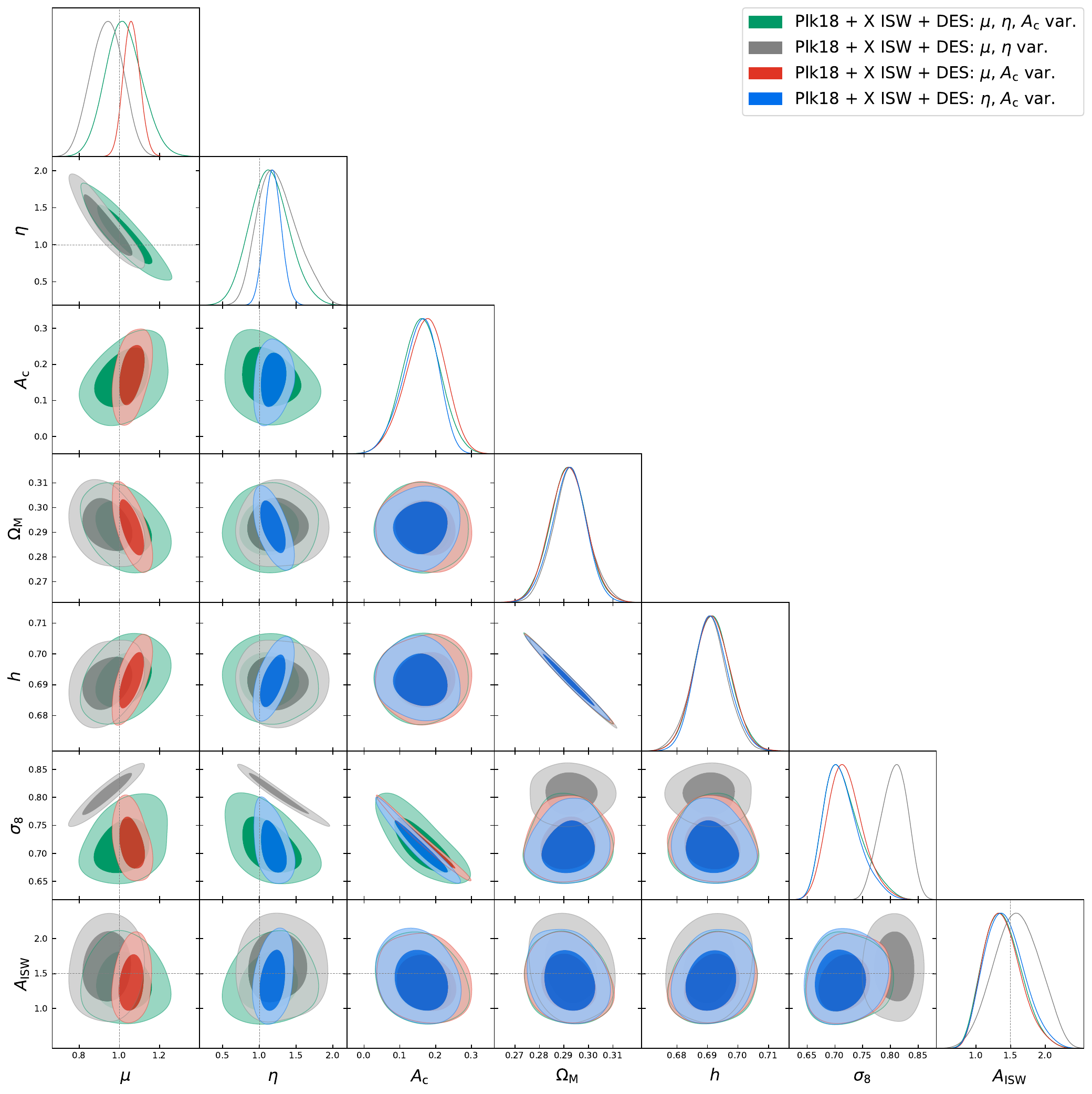}
\caption{Same as Figure~\ref{fig:varprio} but now adding different combinations of CMB $C_{\ell}^{TT,TE,EE}$ and its lensing potential from Plk18. \bigskip}
\label{varCMB}
\end{figure*}

Next, we include also the CMB angular power spectrum. Once more, we start with the constant $A_c,\,\eta,\,\mu$ model, with results shown inin Figure~\ref{cteCMB}, and then consider the time-varying case, which we depict in Figure~\ref{varCMB}.  Adding the CMB is expected to bring further constraints from the late ISW effect, and from the lensing of the CMB along its path. In the constant  case, this has the effect of consolidating the bounds on $\eta$ around the GR value while $\mu$ shows a preference for positive values but is still compatible with GR.  However, $A_{\rm c}$ is positive at above the 2~$\sigma$ level, even when $\mu$ and $\eta$ are also left free. The effect is stronger in the dynamic model than in the constant one.

We also find lower values for $\sigma_8$, close to common local constraints, while the $h$ tension vanishes due to the lower precision. The $A_{\rm c}$-$\eta$ or $A_{\rm c}$-$\mu$ correlations are showing the same trend as the case without the CMB, as expected from the analytic analysis in Sec.~\ref{sec:methanddata} when we include, as necessary in this case, possible effect on both super- and sub-horizon scales. In the dynamical model, we still observe the change in correlation as before, but we see looser constraints now on $A_{\rm c}$ parameter. The only substantial difference is with the $\eta$ parameter now slightly not preferring the unity value.

To summarise and add further insights on the above observed bounds, we say that without including CMB constraints, $\Omega_m$ is correlated with $\sigma_8$ and its impact on the derivative of the power spectrum in the ISW probe dominates a possible correlation between $A_{\rm c}$ and $\sigma_8$, leading $A_{\rm c}$ to still prefer the vanishing value. When $A_{\rm c}$ is allowed to vary with the scale factor, it correlates with $\sigma_8$ and yield a preference for non-null value for $A_{\rm c}$ because ISW and DES prefer a $A_{\rm ISW}$ value far from 1, except when it becomes degenerate with $\mu$ and $\eta$ at which case $\Omega_m$-$\sigma_8$ correlation is seen again. When we combine with CMB, $\Omega_m$ is now strongly constrained, and it loses the correlation with $\sigma_8$ even in the case of $A_{\rm c}$ constant. However, even though that $A_{\rm ISW}$ is compatible with 1, the discrepancy between CMB and the local probes on $\sigma_8$ translates into a deviation of $A_{\rm c}$ from its vanishing value.\footnote{We have checked by running MCMC using ISW+DES alone or combined with CMB from Plk18 that in the latter case the $A_{\rm ISW}$ parameter shifts from $\sim$ 1.57 to one, while $\sigma_8$ is pushed from $\sim$0.78 to $\sim$0.82.}
 
%%%%%%%%%%%%%%%%%%%%%%%%%%%%%%%%%%%%%%%%%%%%%%%%%%%%%%%%%%%%%%%%%%%%%%%%%%%%%%%%%%%%%%%%%%%%%%%%%%%%%%%%%%%%%%%%%%%%%%
\setlength\tabcolsep{2.3pt}
\renewcommand{\arraystretch}{1.8}
\begin{table*}[t]
\centering
\caption{\label{tab:post} 68\%CL marginalized cosmological constraints in constant and dynamical parameterisation for $\mu$, $\eta$ and $A_{\rm c}$ using a variety of datasets. }
%\begin{tabular}{|c|c|c|c|c|c|c|c|c|}
\begin{tabular}{|ccccccccc|}
\hline
%& & & & & & & & \\
Model & Data Sets&$\Omega_{\rm m}$ &$\sigma_8$ &$\mu$ &$\eta$ &$h$ & $A_{\rm c}$ &$A_{\rm ISW}$ \\ 
%& & & & & & & & \\ 
\hline
%& & & & & & & & \\
$A_{\rm c}$ const.\ & CMB$\times$ISW + DES & $0.3_{-0.02}^{+0.02}$ & $0.715_{-0.05}^{+0.035}$ & $\ldots$ & $\ldots$ & $0.61_{-0.046}^{+0.042}$ & $0.49_{-0.29}^{+0.21}$ & $1.014_{-0.4}^{+0.1}$ \\ 
%& & & & & & & & \\ 
$\mu$, $A_{\rm c}$ const.\ & CMB$\times$ISW + DES & $0.305_{-0.2}^{+0.2}$ & $0.78_{-0.053}^{+0.033}$ & $1.3_{-0.27}^{+0.32}$ & $\ldots$ & $0.67_{-0.15}^{+0.15}$ & $0.44_{-0.31}^{+0.27}$ & $1.17_{-0.33}^{+0.079}$ \\ 
%& & & & & & & & \\ 
$\eta$, $A_{\rm c}$ const.\ & CMB$\times$ISW + DES & $0.305_{-0.2}^{+0.2}$ & $0.73_{-0.054}^{+0.033}$ & $\ldots$ & $1.34_{-0.31}^{+0.32}$ & $0.74_{-0.13}^{+0.11}$ & $0.194_{-0.44}^{+0.25}$ & $1.14_{-0.29}^{+0.062}$ \\ 
%& & & & & & & & \\ 
$\mu$, $\eta$, const.\ & CMB$\times$ISW + DES & $0.28_{-0.29}^{+0.29}$ & $0.75_{-0.05}^{+0.038}$ & $1.37_{-0.25}^{+0.25}$ & $1.02_{-0.51}^{+0.52}$ & $0.67_{-0.0049}^{+0.0048}$ & $\ldots$ & $1.29_{-0.41}^{+0.12}$ \\ 
%& & & & & & & & \\ 
$\mu$, $\eta$, $A_{\rm c}$ const.\ & CMB$\times$ISW + DES & $0.28_{-0.30}^{+0.31}$ & $1.3_{-0.054}^{+0.037}$ & $1.07_{-0.28}^{+0.28}$ & $1.05_{-0.51}^{+0.52}$ & $0.825_{-0.0046}^{+0.0044}$ & $0.018_{-0.5}^{+0.5}$ & $1.29_{-0.42}^{+0.13}$ \\ 
%& & & & & & & & \\ 
\hline
%& & & & & & & & \\
$A_{\rm c}$ const.\ & CMB$\times$ISW + DES + CMB &	$0.30_{-0.002}^{+0.002}$	&	$0.76_{-0.028}^{+0.027}$	&	$\ldots$	&	$\ldots$	&	$0.68_{-0.0046}^{+0.0047}$	&	$0.11_{-0.054}^{+0.049}$	&	$1.54_{-0.32}^{+0.29}$	\\ 
%& & & & & & & & \\
$\mu$,  $A_{\rm c}$ const.\ & CMB$\times$ISW + DES + CMB & $0.29_{-0.0012}^{+0.0014}$ & $0.76_{-0.03}^{+0.023}$ & $1.031_{-0.026}^{+0.026}$ & $\ldots$ & $0.69_{-0.0064}^{+0.006}$ & $0.13_{-0.05}^{+0.055}$ & $1.52_{-0.31}^{+0.3}$ \\ 
%& & & & & & & & \\ 
$\eta$, $A_{\rm c}$ const.\ & CMB$\times$ISW + DES + CMB & $0.30_{-0.001}^{+0.0011}$ & $0.76_{-0.033}^{+0.027}$ & $\ldots$ & $1.01_{-0.054}^{+0.045}$ & $0.68_{-0.0049}^{+0.0049}$ & $0.11_{-0.052}^{+0.051}$ & $1.54_{-0.31}^{+0.29}$ \\ 
%& & & & & & & & \\
$\mu$, $\eta$  const.\ & CMB$\times$ISW + DES + CMB & $0.29_{-0.0014}^{+0.0013}$ & $0.82_{-0.014}^{+0.015}$ & $1.01_{-0.027}^{+0.024}$ & $1.025_{-0.053}^{+0.044}$ & $0.69_{-0.0063}^{+0.0058}$ & $\ldots$ & $1.53_{-0.31}^{+0.29}$ \\ 
%& & & & & & & & \\
$\mu$, $\eta$, $A_{\rm c}$ const.\ & CMB$\times$ISW + DES + CMB & $0.29_{-0.003}^{+0.003}$ & $0.75_{-0.034}^{+0.024}$ & $1.03_{-0.026}^{+0.026}$ & $1.02_{-0.054}^{+0.051}$ & $0.69_{-0.0065}^{+0.0058}$ & $0.13_{-0.049}^{+0.053}$ & $1.51_{-0.31}^{+0.28}$ \\ %& & & & & & & & \\
\hline
%& & & & & & & & \\
$A_{\rm c}(a)$\ & CMB$\times$ISW + DES &	$0.30_{-0.02}^{+0.02}$	&	$0.70_{-0.061}^{+0.038}$	&	$\ldots$	&	$\ldots$	&	$0.64_{-0.049}^{+0.041}$	&	$0.27_{-0.2}^{+0.14}$	&	$1.10_{-0.26}^{+0.042}$	\\ %& & & & & & & & \\
$\mu(a)$, $A_{\rm c}(a)$\ & CMB$\times$ISW + DES & $0.22_{-0.21}^{+0.22}$ & $0.66_{-0.059}^{+0.026}$ & $1.54_{-0.2}^{+0.35}$ & $\ldots$ & $0.85_{-0.14}^{+0.13}$ & $0.68_{-0.087}^{+0.32}$ & $1.26_{-0.41}^{+0.1}$ \\ %& & & & & & & & \\ 
$\eta(a)$, $A_{\rm c}(a)$\ & CMB$\times$ISW + DES & $0.25_{-0.22}^{+0.21}$ & $0.665_{-0.062}^{+0.031}$ & $\ldots$ & $1.59_{-0.11}^{+0.42}$ & $0.74_{-0.088}^{+0.087}$ & $0.43_{-0.25}^{+0.18}$ & $1.14_{-0.29}^{+0.05}$ \\ %& & & & & & & & \\ 
$\mu(a)$, $\eta(a)$ \ & CMB$\times$ISW + DES & $0.305_{-0.031}^{+0.032}$ & $0.76_{-0.049}^{+0.035}$ & $1.103_{-0.45}^{+0.21}$ & $0.92_{-0.45}^{+0.21}$ & $0.64_{-0.098}^{+0.087}$ & $\ldots$ & $1.37_{-0.53}^{+0.15}$ \\ %& & & & & & & & \\ 
$\mu(a)$, $\eta(a)$, $A_{\rm c}(a)$\ & CMB$\times$ISW + DES & $0.305_{-0.031}^{+0.031}$ & $0.76_{-0.05}^{+0.036}$ & $1.10_{-0.43}^{+0.22}$ & $0.90_{-0.45}^{+0.51}$ & $0.65_{-0.095}^{+0.085}$ & $0.015_{-0.3}^{+0.3}$ & $1.39_{-0.54}^{+0.16}$ \\ %& & & & & & & & \\ 
\hline
%& & & & & & & & \\
$A_{\rm c}(a)$\ & CMB$\times$ISW + DES + CMB &	$0.30_{-0.002}^{+0.002}$	&	$0.74_{-0.039}^{+0.03}$	&	$\ldots$	&	$\ldots$	&	$0.69_{-0.0046}^{+0.005}$	&	$0.137_{-0.055}^{+0.058}$	&	$1.32_{-0.28}^{+0.24}$	\\ %& & & & & & & & \\
$\mu(a)$, $A_{\rm c}(a)$\ & CMB$\times$ISW + DES + CMB & $0.29_{-0.0011}^{+0.0013}$ & $0.72_{-0.037}^{+0.025}$ & $1.15_{-0.058}^{+0.059}$ & $\ldots$ & $0.69_{-0.0063}^{+0.0057}$ & $0.17_{-0.051}^{+0.061}$ & $1.38_{-0.31}^{+0.23}$ \\ %& & & & & & & & \\ 
$\eta(a)$, $A_{\rm c}(a)$\ & CMB$\times$ISW + DES + CMB & $0.29_{-0.0014}^{+0.0011}$ & $0.71_{-0.038}^{+0.022}$ & $\ldots$ & $1.18_{-0.12}^{+0.11}$ & $0.69_{-0.0062}^{+0.005}$ & $0.157_{-0.042}^{+0.056}$ & $1.41_{-0.33}^{+0.25}$ \\ %& & & & & & & & \\
$\mu(a)$, $\eta(a)$ \ & CMB$\times$ISW + DES + CMB & $0.29_{-0.0013}^{+0.0011}$ & $0.81_{-0.022}^{+0.028}$ & $0.90_{-0.43}^{+0.43}$ & $1.24_{-0.32}^{+0.22}$ & $0.69_{-0.0055}^{+0.0059}$ & $\ldots$ & $1.59_{-0.35}^{+0.34}$ \\ %& & & & & & & & \\
$\mu(a)$, $\eta(a)$, $A_{\rm c}(a)$\ & CMB$\times$ISW + DES + CMB & $0.29_{-0.003}^{+0.003}$ & $0.72_{-0.041}^{+0.024}$ & $1.06_{-0.52}^{+0.52}$ & $1.26_{-0.23}^{+0.23}$ & $0.69_{-0.0063}^{+0.0054}$ & $0.16_{-0.054}^{+0.054}$ & $1.38_{-0.32}^{+0.24}$ \\ %& & & & & & & & \\ 
\hline
\end{tabular}
\bigskip
\end{table*}

\bigskip

\section{Conclusion}\label{conclusion}

In this work, we wanted to test the conservation of the continuity equation at the perturbation level by means of a phenomenological modelling of its deviation from the standard behaviour. We parameterise this with a simple $A_{\rm c}$ parameter, which is still however compatible with allowed extensions from general theoretical considerations. We also test the interplay of the continuity deviation parameter with two other common parameters $\mu$ and $\eta$ relative to the functions modelling the deviation from general relativity of the Poisson and the Weyl potential. For that we performed a Bayesian analysis using different combinations of the CMB temperature, polarization and lensing potential angular spectrum, along with the ISW galaxy cross power spectrum and the 3$\times$2pt lensing and galaxy clustering from DES survey probes. This choice was motivated by the fact that $\mu$ mainly affects the growth of structure especially on sub horizon scales while $\eta$ impacts more the lensing, or the ISW effect from the LSS, or to a very lesser degree the perturbations at super horizon scales in probes such as CMB angular temperature and polarization spectra. To close our system of equations, we supposed two functional parametrisation of the studied parameters, a constant or a dynamical late time variation.

Without combining with the CMB datasets, we found that our deviation continuity parameter, in the constant or dynamical parameterisation, is still compatible with $\Lambda$CDM, though showing a preference of 2--3$\sigma$ for non-null values for some cases, either alone or when we additionally allow either $\mu$ or $\eta$ to vary. We also observed for most of the combinations, as expected from our prior theoretical investigation, a strong correlation between the Poisson or anisotropic stress and the continuity parameter. When we add CMB constraints, we found the same behaviour as before for either of the parameterisations with moreover a clear 2$\sigma$ deviation from the null value for $A_{\rm c}$ in almost all the cases. These deviations from $\Lambda$CDM are mainly driven by, either discrepancy on the $A_{\rm ISW}$ values or on $\sigma_8$ ones between CMB and local probes. This calls for taking the assumption of the conservation of the continuity equation with care, especially when testing additional deviations from general relativity at the perturbation level with growth of structure probes.

\bigskip
\section*{Acknowledgements}
We thank Luca Amendola for several discussions and for inputs on the early drafts. ZS acknowledges support from DFG project 456622116 during which this work was conducted and support from the research projects PID2021-123012NB-C43, PID2024-159420NB-C43, the Proyecto de Investigación SAFE25003 from the Consejo Superior de Investigaciones Científicas (CSIC), and the Spanish Research Agency (Agencia Estatal de Investigaci\'on) through the Grant IFT Centro de Excelencia Severo Ochoa No CEX2020-001007-S, funded by MCIN/AEI/10.13039/501100011033.  MQ is supported by the Brazilian research agencies FAPERJ project E-26/201.237/2022, CNPq and CAPES.


\bibliographystyle{apsrev4-2.bst}

%\pagebreak

\bibliography{Ref.bib}

\end{document}